\documentstyle[12pt]{article}

\begin{document}


\def\a{\alpha}
\def\b{\beta}
\def\c{\varepsilon}
\def\d{\delta}
\def\e{\epsilon}
\def\f{\phi}
\def\g{\gamma}
\def\h{\theta}
\def\k{\kappa}
\def\l{\lambda}
\def\m{\mu}
\def\n{\nu}
\def\p{\psi}
\def\q{\partial}
\def\r{\rho}
\def\s{\sigma}
\def\t{\tau}
\def\u{\upsilon}
\def\v{\varphi}
\def\w{\omega}
\def\x{\xi}
\def\y{\eta}
\def\z{\zeta}
\def\D{\Delta}
\def\G{\Gamma}
\def\H{\Theta}
\def\L{\Lambda}
\def\F{\Phi}
\def\P{\Psi}
\def\S{\Sigma}

\def\o{\over}
\def\beq{\begin{eqnarray}}
\def\eeq{\end{eqnarray}}
\newcommand{\gsim}{ \mathop{}_{\textstyle \sim}^{\textstyle >} }
\newcommand{\lsim}{ \mathop{}_{\textstyle \sim}^{\textstyle <} }

\def\IJMP{Int.~J.~Mod.~Phys. }
\def\MPL{Mod.~Phys.~Lett. }
\def\NP{Nucl.~Phys. }
\def\PL{Phys.~Lett. }
\def\PR{Phys.~Rev. }
\def\PRL{Phys.~Rev.~Lett. }
\def\PTP{Prog.~Theor.~Phys. }
\def\ZP{Z.~Phys. }


\baselineskip 0.7cm

\begin{titlepage}
\begin{flushright}
UT-964
\\
August, 2001
\end{flushright}

\vskip 1.35cm
\begin{center}
{\large \bf
Tubular Field Theory
}
\vskip 1.2cm
Izawa K.-I.
\vskip 0.4cm

{\it Department of Physics and RESCEU, University of Tokyo,\\
     Tokyo 113-0033, Japan}

\vskip 1.5cm

\abstract{
We propose tubular field theory,
which is a continuum analogue of lattice field theory.
One-dimensional links (and zero-dimensional sites)
in lattice field theory
are replaced by two-dimensional tubes
to result in two-dimensional spacetime microscopically.
As an example, scalar and gauge fields are considered
in `three-dimensional' tubular spacetime.
}
\end{center}
\end{titlepage}

\setcounter{page}{2}


It is conceivable that quantum gravity effects
yield microscopic fluctuations of spacetime 
such as spacetime foam at small scales.
The observed four-dimensional spacetime might be
macroscopic  coarse-graining of such microscopic structures.

In this paper, we consider two-dimensional surface
as a simplest tractable model of the microscopic
structure and macroscopic spacetime manifold
to be large-scale approximation thereof.
Although the motivation is closely related to gravity,
we restrict ourselves to non-gravitational
(and `three-dimensional') setup as a first step.

For concreteness, let us imagine the three-dimensional
cubic lattice with its one-dimensional links replaced by 
two-dimensional cubic tubes
of the size $l$ (the area of the cubic tube $T$
with topology $S^1 \times [0, l]$ amounts to $4l^2$
since the circumference of the square $S^1$ is $4l$).
At the `site' $(i, j, k) \in {\bf Z}^3$,
six cubic tubes are joined
with two tubes extended in the two opposite directions along each axis
$x, y$, or $z$.
These tubes constitute a two-dimensional surface
$\S$ as a whole and microscopically the spacetime is two-dimensional.

As a first example, we consider a scalar field on the surface $\S$
with the action
\beq
 S = \int_\S \! d^2\x \, \left({1 \o 2}\q_a \f \q^a \f
   - V(\f)\right), 
\eeq
where $a=1,2$ denotes a two-dimensional index.

At large scales,
dropping the Kaluza-Klein modes
of $\f$ in the circular direction $S^1$ of the tube
$T \sim S^1 \times [0, l]$,
we obtain
\beq
 S \simeq \sum_{(i, j, k)} \sum_\mu
   \int_{T_\mu} \! d^2\x \, \left({1 \o 2}
   \q_\mu \f \q^\mu \f - V(\f)\right), 
\eeq
where $\mu = x,y,z$ represents the positive direction along each axis
perpendicular to the circular direction
and $T_\mu$ denotes the corresponding tube
at the `site' $(i, j, k)$.
Further approximation goes as follows {\` a} la lattice field theory:
\beq
 S \simeq 4l^2 \sum_{(i, j, k)} \left({1 \o 2}
   \q_\mu \f \q^\mu \f - 3V(\f)\right)
   \simeq \int \! dxdydz \, \left({1 \o 2}
   \q_\mu {\tilde \f} \q^\mu {\tilde \f} - {\tilde V}({\tilde \f})
   \right),
\eeq
where we have defined
\beq
 {\tilde \f} = {1 \o \sqrt{2l}}\f, \quad
 {\tilde V}({\tilde \f}) = {3 \o 2l}V(\f).
\eeq
In fact, coarse-graining through path integration
affects the form of ${\tilde V}({\tilde \f})$
and relevant terms in it should become dominant.

As a next example, we consider (nonabelian) gauge theory on $\S$
with gauge field $A_a$ and charged scalar $\f$.
The gauge-invariant action is given by
\beq
 S = \int_\S \! d^2\x \left({1 \o 4}F_{ab}F^{ab}
   + D_a \f (D^a \f)^* - V(\f)\right).
\eeq

Let us adopt an axial gauge where the gauge field component
along the circular direction is vanishing $A_{S^1} = 0$.
Dropping the Kaluza-Klein modes, as is the case for the scalar
field above, we obtain
\beq
 S \simeq \sum_{(i, j, k)} \sum_\mu
   \int_{T_\mu} \! d^2\x \, \left(
   D_\mu \f (D^\mu \f)^* - V(\f)\right).
\eeq
Further approximation goes as follows:
\beq
 S \simeq 4l^2 \sum_{(i, j, k)} \left(
   D_\mu \f (D^\mu \f)^* - 3V(\f)\right)
   \simeq \int \! dxdydz \, \left(
   D_\mu {\tilde \f} (D^\mu {\tilde \f})^* - {\tilde V}({\tilde \f})
   \right),
\eeq
where we have defined
\beq
 {\tilde \f} = {1 \o \sqrt{2l}}\f, \quad
 {\tilde V}({\tilde \f}) = {3 \o 2l}V(\f).
\eeq
In this case, coarse-graining through path integration
should generate the kinetic term
for the gauge field $A_\mu$
with gauge invariance, which remains after `higher-dimensional'
gauge fixing $A_{S^1} = 0$.

The above two examples imply that three-dimensional
(and more generally, higher-dimensional) field theories
may arise as large-scale approximations even though microscopically
the underlying spacetime is two-dimensional
(or more generally, lower-dimensional).
To make this approach more realistic, it seems necessary
to include (chiral) fermions and gravity in the present framework.
In particular, introduction of gravity may naturally make effective dimensions
of the macroscopic spacetime to be variable, which should be
determined dynamically. Since the underlying spacetime is
two-dimensional rather than one-dimensional, such higher-dimensional
geometry may be accommodated intrinsically.

\section*{Acknowledgments}

We would like to thank T.~Watari and T.~Yanagida
for valuable discussions.

\end{document}